\documentclass[aps,prl,twocolumn,superscriptaddress]{revtex4-2}
\usepackage{dcolumn}
\usepackage{bm}
\usepackage{graphicx}
\usepackage{ulem}
 \usepackage{amssymb,wasysym}
 \usepackage{xcolor}
 
\usepackage[colorlinks=true, allcolors=red]{hyperref}
\DeclareTextSymbol{\degre}{OT1}{23}

\begin{document}
\title{Strengthening by softening: rigidity increase of a curved sheet from  nonlinear regime of deformation }
\author{Nino Quillent-Elinguel and Thomas Barois}
\affiliation{Univ. Bordeaux, CNRS, LOMA, UMR 5798, F-33400 Talence, France}
\begin{abstract}
It is well-known that a thin sheet held in a rigid circular clamp has a larger flexural strength than when it is flat. Here, we report that the flexural strength of curved sheets is further increased with a softening of the clamping condition. 
This unexpected compliance effect relates to the geometrical properties of curvature-induced rigidity that we observe in controlled experiments and further analyze with numerical simulations. In addition, we identify another compliance effect in which opened curved sheets can be more resistant to bending than closed cylinders of same dimensions.   
\end{abstract}
\maketitle

Thin sheets are mechanical structures presenting a rich phenomenology\cite{audoly2005fragmentation,marder2007crumpling,audoly2010elasticity,vandeparre2011wrinkling,reis2015perspective,holmes2019elasticity,croll2019compressive,siefert2021stretch} due to the geometrical effects arising from the large mismatch between  small thicknesses and comparatively large sizes. For a rectangular sheet of length $L$, width $W$ and thickness $t$, the scaling relations $t \ll W$ and $t \ll L$ make that the sheet naturally favors bending deformations over stretching deformations\cite{rayleigh1888bending}. A consequence of this is the possibility to induce stiffening of floppy sheets\cite{pini2016two}: by imposing a curvature in a given direction, the bending in the other direction becomes energetically expensive because it requires  stretching. Curvature-induced rigidity is found in different contexts such as tape springs\cite{calladine2020folding}, folded strips\cite{walker2019flexural}, plant leaves\cite{moulia1994mechanics,moulia2000leaves,wei2023unveiling}, egg geometry\cite{lazarus2012geometry}, fins\cite{nguyen2017curvature} and wings\cite{macphee2022aerodynamic} stiffening, actuators\cite{wu2021printed}, wavy walls\cite{o1980walls} and pizza slice manipulation\cite{taffetani2019limitations}.

When thin sheets are curved, the structural failure\cite{peterson1968buckling,singer2002vol} under loads usually follows a different phenomenology than the bulk material failure. The failure of shells under bending or compression loads is often not a failure of the material itself but the result of the growth of a buckling mode predicted by a linear stability analysis\cite{houliara2010stability,hutchinson2010knockdown,rotter2014nonlinear,virot2017stability,gerasimidis2018establishing,gerasimidis2020dent}. 
In the post-buckling regime, curved sheets
exhibit some form of crumpling\cite{ben1997crumpled,witten2007stress}, which takes the form of a more or less complex network of singularities\cite{cerda1998conical,chaieb1998experimental,cerda1999conical,gottesman2015furrows,chopin2016disclinations,audoly2020localization} and ridges\cite{lobkovsky1996boundary,didonna2002scaling,venkataramani2003lower,andresen2007ridge,cambou2011three,fuentealba2015transition}. The formation of this type of structures was studied for compaction\cite{blair2005geometry,vliegenthart2006forced,sultan2006statistics,balankin2010fractal,nasto2013localization,gottesman2018state,abramian2020nondestructive,bense2021complex}, indentation\cite{boudaoud2000dynamics,vaziri2008localized,vella2012indentation} or in other situations\cite{roman2012stress,barois2021transition}.

\begin{figure}[h!]
    \centering
    \includegraphics[width=8.5cm]{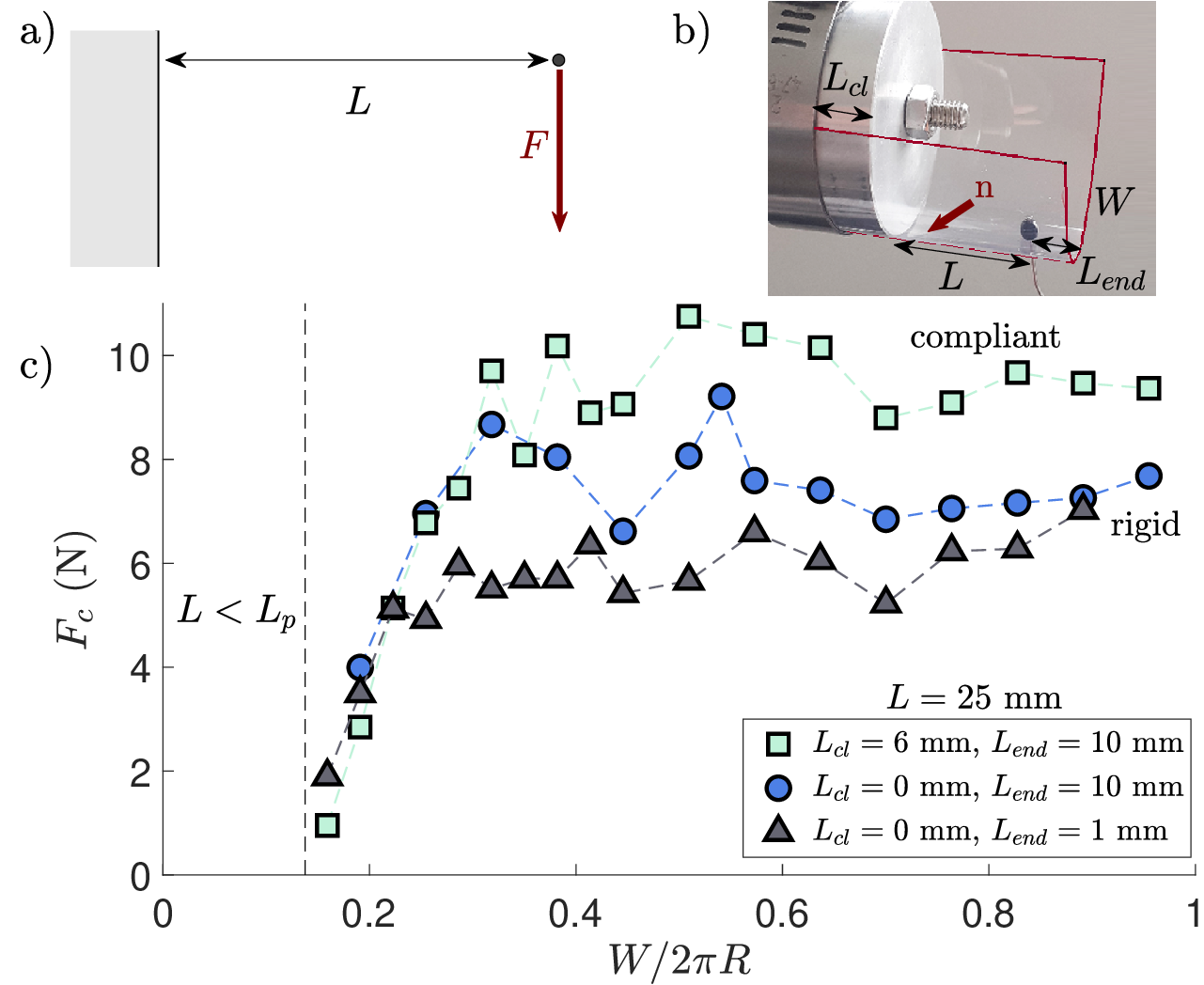}
    \caption{a) Question addressed in this work: what is the optimal structure to support a force $F$ a distance $L$? b) Rectangular sheet of width $W$ and thickness $t$ held in a circular clamp of radius $R=25$ mm. $L_{cl}$ is the distance between the inner circular cylinder and the external part of the clamp. $L=25$ mm is the distance between the clamp and the applied force. $L_{end}$ is the distance between the applied force and the free end of the sheet. $n$ points to the location of the stress focusing singularities for collapsed sheets. c) 
    Collapse forces as a function of the normalized width   $W/2\pi R$  that quantifies the portion of the circular clamp covered by the sheet. The forces are applied by masses hanging under gravity. $L_{cl} = 0$ mm corresponds to a rigid clamp and $L_{cl} = 6$ mm is for a compliant clamp. }
    \label{fig:exp}
\end{figure}

The question we address is this work is illustrated in figure \ref{fig:exp}a: what is the optimal structure that can support a static load $F$ at a horizontal distance $L$ of a rigid wall? With the classical Euler-Bernoulli beam theory as a starting approach, it seems that, first, there is no need for the structure to expand beyond the distance $L$ and, second, the anchoring of an optimal structure with the rigid wall should be as rigid as possible. We will show that these two intuitive ideas are wrong for model structures in the non-linear regime of curvature-induced rigidity.

We consider a simple mechanical structure obtained by curving a piece of rectangular thin sheet, as shown in figure \ref{fig:exp}b. 
The sheets are initially flat and rectangular with thickness $t=0.2$ mm and width $W$. The circular clamp is a rigid cylinder of radius $R = 25$ mm and the  curved sheet configuration is obtained by pressing the sheet against the rigid cylinder with a hose clamp. The distance $L_{cl}$ denotes the shift in position between the hose clamp and the cylinder. $L_{cl}=0$ mm is the rigid clamping condition for which the sheet is forced to lie on a circle at the end of the clamp. $L_{cl} >0$ mm is a soft clamp condition in which the sheet is circular without applied load but in the presence of loads, the sheet can slightly move away from the inner cylinder. The free-standing length of the sheet is $L + L_{end}$ where $L_{end}$ is the distance between the position of the applied force and the free edge of the sheet. The distance $L$ is the effective length between the end of the clamp and the location of the applied force. The load is applied by hanging a mass to a hook passing through a small hole. 
For long sheets, the persistence length\cite{lobkovsky1997properties,mahadevan2007persistence,barois2014curved,matsumoto2018pinching,bhaskar2021far,audoly2021one} of the transverse curvature scales as $L_p \sim W^2/\sqrt{Rt}$. Here, we are interested in the regime of curvature-induced stiffening $L<L_p$, which means that, in the absence of applied forces, the sheets are sufficiently short for the transverse curvature to be strictly positive for the whole sheet.

In figure \ref{fig:exp}c, we represent the measured collapse force $F_c$ as a function of the width normalized by the perimeter $W/2\pi R$ for 3 sets of data. 
The reference set of data with triangle symbols is for the force applied through a hole placed as close as possible to the end of the sheet ($L_{end}= 1$ mm) and a rigid clamp $L_{cl} = 0$ mm.  The first data point is for $W/2\pi R$ = 0.16. Below this value, the sheet is floppy because $W$ is not large enough for the sheet to be curved at a distance $L$. The region $L<L_p$ is defined by the persistence length\cite{barois2014curved} is $L_p = W^2/\sqrt{70 R t}$ or equivalently $W/2\pi R = {L_p}^{1/2}(70 R t)^{1/4}/2\pi R = 0.138$ with $L_p = 25$ mm, $R = 25$ mm and $t = 0.2$ mm.
In the region $W/2\pi R \sim 0.2$, the sheet rigidity increases and reaches a plateau at about 6 N for the reference sheet with rigid clamping ($L_{cl} = 0$ mm) and the force applied as close as possible to the end of the sheet ($L_{end} = 1$ mm). The two other data sets are for the same length $L$ but with a longer  sheet $L+L_{end}$ with $L_{end} = 10$ mm. For a rigid clamping $L_{cl} = 0$ mm, the plateau is higher than for $L_{end}=1$ mm. This is a first surprising result because the length $L$ involved in the bending moment was unmodified. In the classical Euler-Bernoulli beam theory, the mechanical stiffness of beams is local. Here, the modification of $L_{end}$ has an effect on the buckling nucleation process that takes place close to the cylinder (indicated by $n$ in the figure). The last set of data with square symbol is the compliant clamping obtained by moving backwards the clamping hose of a distance $L_{cl} = 6$ mm without changing any other parameters. We find that the collapse force $F_c$ is even larger. Again, this increased rigidity is surprising here because the bending length was increased from $L = 25$ mm to $L+L_{cl}=31$ mm.

\begin{figure}[h!]
    \centering
    \includegraphics[width=8.5cm]{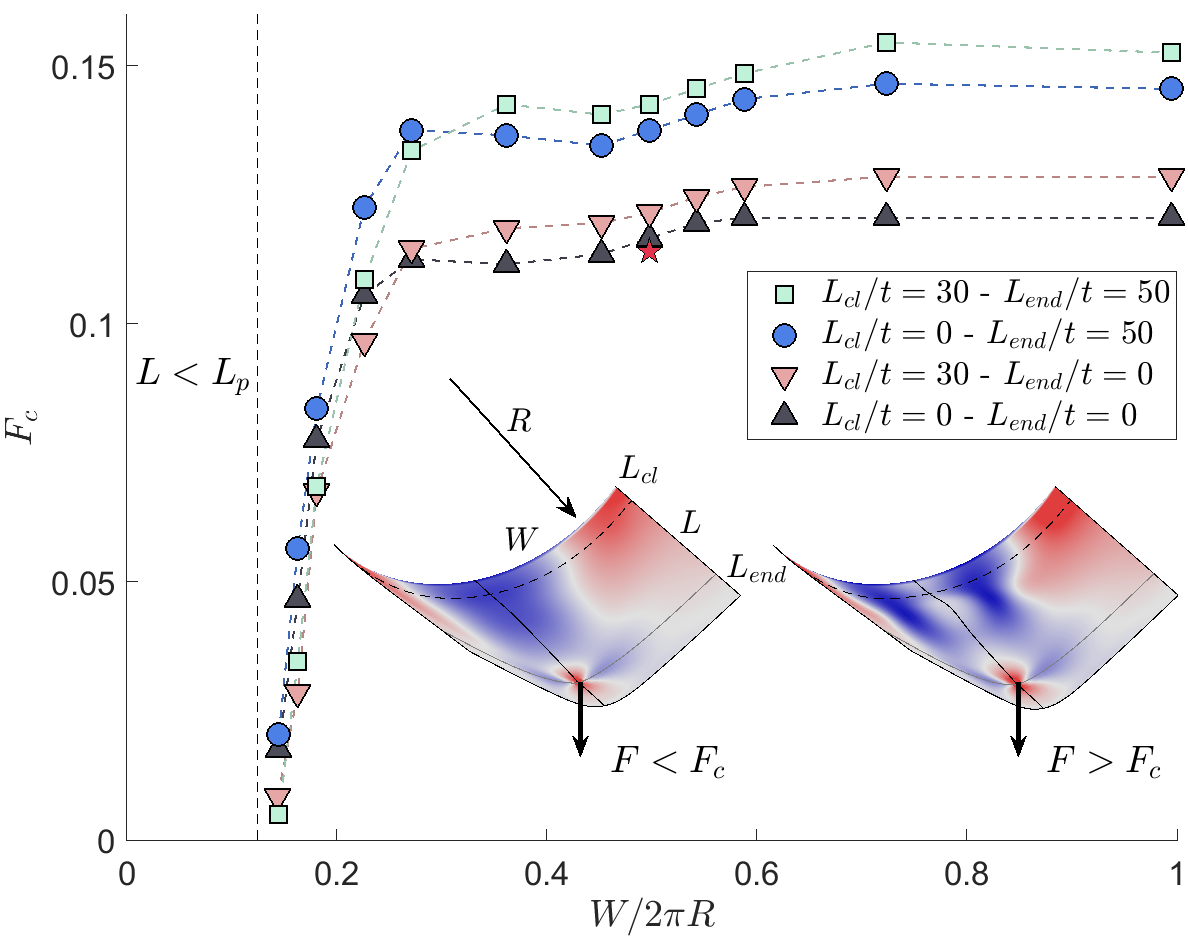}
    \caption{Spring model simulation for the collapse forces $F_c$ for rectangular sheets of length $L = 170 \times t$ held in a circular clamped of radius $R = 175 \times t$. The width is varied between $W = 160 \times t$ ($W/2\pi R \approx 0.145$) and $W = 1100\times t$ ($W/2\pi R \approx 1$, full circle covering at the clamp).   Two simulated sheets ($W/2\pi R = 0.3$) are represented with $F$ below and above the collapse force $F_c$. The collapse is characterized by a splitting of the compression pattern in two locations symmetrically to the longitudinal centerline. A buckle of the centerline is also visible in the collapsed sheet $F>F_c$.   }
    \label{fig:simu_W}
\end{figure}

To analyze in more details the peculiar regimes observed in experiments, we numerically solved a spring-network model\cite{barois2021transition} of curved sheets under loads (see movie M1 for an animation of the simulation results of a collapse event and the supplemental material\cite{supplemental} for a description of the model). In figure \ref{fig:simu_W}, we compute the critical force $F_c$ to collapse a sheet of reduced length $L/t = 170$ maintained on a circular clamp of radius $R/t = 175$. The varied parameter in the simulation are $W$, $L_{cl}$ and $L_{end}$ but with fixed $L$. $W$ is normalized by the clamp perimeter  $  W/2\pi R$. We identify the 3 regimes found in the experiments with a vanishing $F_c$ for small $W/2\pi R$, a rapid increase for $W/2\pi R$ of the order of 0.2 and a plateau reached at $W/2\pi R\sim 0.3$. As in the experiments, we find that an increase of $L_{end}$ and an increase of $L_{cl}$ results in a larger collapse force in the plateau region $W/2\pi R > 0.3$. 

The configuration of the sheet obtained by simulation of the spring model is represented in figure  \ref{fig:simu_W} for $W/2\pi R = 0.27$ with $L_{cl}/t = 30$ and $L_{end}/t = 50$ for $F<F_c$ (no collapse) and $F>F_c$ (collapsed). The color contour represented on the sheet's surface is the value of the longitudinal strain with the hot regions corresponding to positive stretching and the cold regions corresponding to the regions under compression. For a non-collapsed sheet, the axial compression is maximal on the sheet centerline. The collapse for $F>F_c$ corresponds to a buckling event with two stress-focusing singularities appearing symmetrically of the centerline and connected by a small ridge that is visible by a bump forming on the centerline.

From the results of the simulation of the sheet profile, we can compute the sheet's transverse curvature on the centerline. The local curvature of the sheet is an important quantity because the buckling instability under global bending is ruled by a mechanical stability criterion\cite{hutchinson2010knockdown}: 
\begin{equation}
    \sigma_c(u) \sim -E \frac{t}{R(u)}
\end{equation}
in which $E$ is the Young modulus, $t$ is the sheet thickness, $c(u) = 1/R(u)$ is the sheet curvature at position $u$ and $\sigma_c(u)$ the critical in-plane stress. The minus sign stipulates that the buckling occurs in the regions of compression. 

\begin{figure}[h!]
    \centering
    \includegraphics[width=8.5cm]{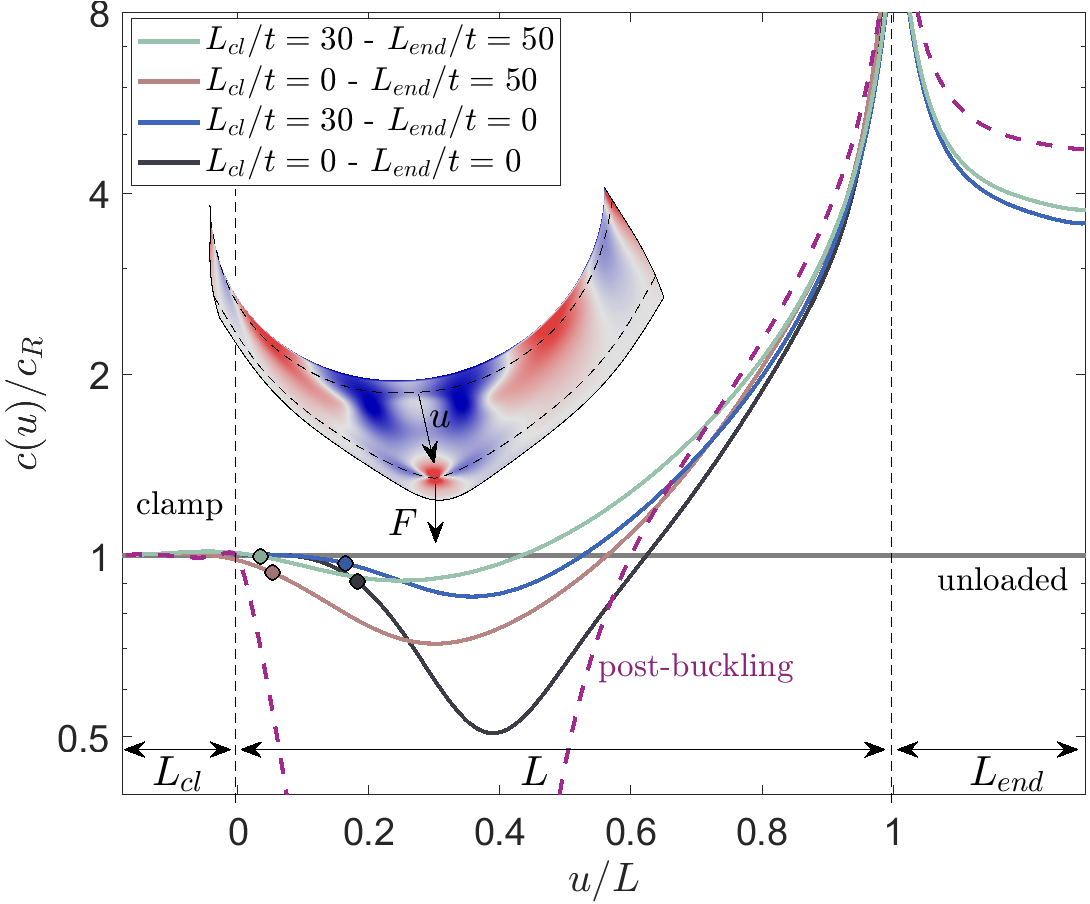}
    \caption{Transverse curvature $c(u)$ in the middle of the sheet as a function of the normalized distance to the clamp $u/L$ for different values of the boundary conditions $L_{end}/t$ and $L_{cl}/t$ with $F=0.12$ (uncollapsed sheet $F<F_c$). The curvature is normalized by the imposed curvature $c_R = R^{-1}$. The point force $F$ applies at $u/L = 1$. The clamp radius is $R=175\times t$ and the length of the sheet is $L = 170 \times t$.  The dashed line is the post-buckling curvature for $L_{cl}/t = 30$ and $L_{end}/t=50$. A collapsed sheet is represented in the figure with the variable $u$ and the color indicates the longitudinal strain (hot: positive stretching, cold: compression). }
    \label{fig:curv}
\end{figure}

Figure \ref{fig:curv} represents the transverse curvature $c(u)$ on the centerline of the curved sheet for the 4 boundary conditions with $L_{end}/t$ equals to 0 or 50 and $L_{cl}/t$ equals 0 or  30 and the same $L/t= 170$. The curvature is normalized by the curvature $c_R = 1/R$ imposed by the circular clamp and it is represented in dimensionless unit $u/L$. $u/L=0$ is the position of the end of the inner cylinder. For $L_{cl}/t= 0$, $u/L$ is defined in a negative range. $u/L= 1$ is the position of the applied force $F$. The 4 plots on figure \ref{fig:curv} are for a half-circular sheet $W/2\pi R = 0.5$ and a force $F = 0.116$ below the collapse value $F_c$. Without applied force $F = 0$ (unloaded), the dimensionless curvature $c(u)/c_R$ remains close to 1. For $u/L \sim 1$, $c(u)/c_R$ gets very large because the force is applied locally on the sheet and forms a cusp. 

The important point in figure \ref{fig:curv} is that the local curvature $c(u)/c_R$ is larger than its unloaded value when a force $F$ is applied. Because the bending resistance of the sheet increases with the transverse curvature $c(u)$, there is a competition between the global bending moment that tends to flatten the sheet in the region $u/L \sim 0.3$ and the effect of the load that locally deforms the sheet by increasing its transverse curvature. Either when $L_{end}/t=50$ and/or $L_{cl}/t=30$, we can notice that the curvature decays on a longer scale than for the reference $L_{end}/t=0$, $L_{cl}/t=0$. Concerning the clamping condition with $L_{cl}/t$, we can say that the effect of the soft clamp $L_{cl}/t = 30$ is to let the sheet have a more pronounced curvature by a release in the absence of the external part of the clamp. 

\begin{figure}[h!]
    \centering
    \includegraphics[width=8.5cm]{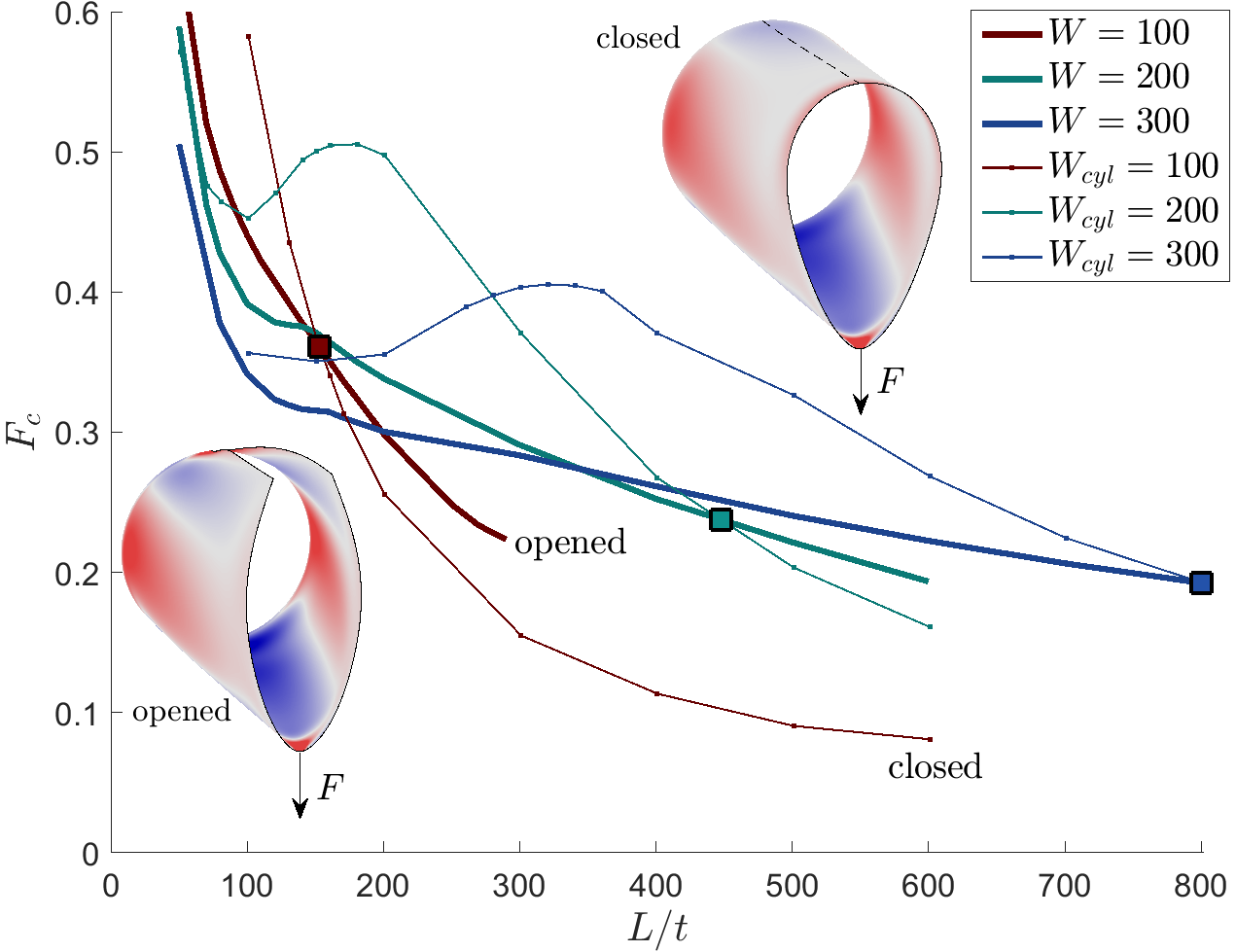}
    \caption{Simulated collapse forces $F_c$ for opened and closed circular sheets with the same clamping conditions with $R = W/2\pi$. $F_c$ is represented as a function of the sheet's normalized length $L/t$.  For the 3 sets of tests with $W/t = 100$, 200 and 300, a square symbol identifies
    the crossover between stronger cylinders at small $L/t$ and stronger opened sheets at large $L/t$. $W_{cyl}$ is the perimeter of the closed sheets.}
    \label{fig:open_vs_closed}
\end{figure}

In the last part of this paper, we would like to provide another illustration of enhanced curvature-induced rigidity by effect of compliance. In figure \ref{fig:open_vs_closed}, we show the simulation results for the collapse forces for two type of geometries: first, close circular sheets with perimeter $W_{cyl}$ in circular clamp and, second, opened sheets of width $W$ with the clamping conditions for $W/2\pi R = 1$.
 The figure represents the collapse force $F_c$ as a function of the dimensionless length $L/t$ in the case $L_{end}/t=0$ (applied force at the end point of the curved sheet) and $L_{cl}/t = 0$ (rigid clamping condition). For the 3 sets of data for $W/t = 100$, 200 and 300, we find that the closed cylinders have larger collapse forces for small $L/t$ but a crossover is observed at $L/t = 153$, 447 and 800 respectively in which the opened sheet configurations can support larger loads than the equivalent closed sheets. In a practical situation, this means that a clamped long cylinder can be made more resistant under bending if a longitudinal cut is made on the whole length of the cylinder, which is reminiscent of the mechanical version of the Braess paradox\cite{cohen1991paradoxical}. The origin of this enhanced rigidity for opened curved sheets is again related to curvature-induced rigidity. In supplemental material\cite{supplemental}, we plot the transverse curvature $c(u)$ for the closed cylinder and the opened curved sheet and we find that the opening of the curved sheet goes against the flattening of the cylinder at the origin of buckling under global bending. 

In this letter, we showed that thin sheets held curved in circular clamps of radius $R$ can show unexpected strengthening regimes under bending. For short sheets with $L\sim R$, we found both in experiments and simulations that a soft clamps can lead to  more resistant structures under global bending. The softness of the clamp is controlled by a length $L_{cl}$ that measures the shift between the internal and the external parts of the circular clamp. We also showed that having a sheet longer than position of the applied force by a quantity $L_{end}$ also leads to a stronger structure. We also found in simulations that for long sheets $L > R$, a regime of stronger opened sheets exists for which opened curved sheets have a larger bending strength than their equivalent closed cylinder. 

The origin of the surprising regimes reported in this paper comes from two competing effects associated to the applied bending force. On the one hand the bending force induces a longitudinal compression stress at the bottom of the curved sheet that favors the buckling of the curved sheet but on the other hand, the point force induces a deformation of the curved sheet and reinforces the sheet by locally increasing its transverse curvature.  These results are of a significant importance for the design and the understanding of the resilience of soft minimal structures.   

\bibliography{biblio}
\end{document}